\documentclass[12pt,preprint]{aastex}
\usepackage{natbib}
\usepackage{lscape}

\newcommand{\hnc}{\mbox{\rmfamily HNC}\,{(1--0)}\,}

\newcommand{\kms}{\mbox{\,km\,s$^{-1}$}}
\newcommand{\msun}{\,$M_{\odot}$}
\newcommand{\degree}{$^{\circ}$}

\newcommand{\HII}{\mbox{$\mathrm{H\,{\scriptstyle {II}}}$}\,}

\shorttitle{The Nessie Nebula}
\shortauthors{Jackson et al.}

\begin{document}

\title{The ``Nessie'' Nebula: Cluster Formation in a Filamentary Infrared Dark Cloud}

\author{James M. Jackson, Susanna C. Finn}
\affil{Institute for Astrophysical Research, Boston University, Boston, MA 02215; jackson@bu.edu, sfinn@bu.edu}

\author{Edward T. Chambers}
\affil{Department of Physics and Astronomy, Northwestern University, Evanston, IL 60208; e-chambers@northwestern.edu}

\author{Jill M. Rathborne}
\affil{Australia Telescope National Facility and Universidad de Chile, Santiago, Chile; rathborn@das.uchile.cl}
\and
\author{Robert Simon}
\affil{I.Physikalisches Institut, Universit\"at zu K\"oln, 50937 K\"oln, Germany; simonr@ph1.uni-koeln.de}

\begin{abstract}

The ``Nessie" Nebula is a filamentary infrared dark cloud (IRDC) with a
large aspect ratio of over 150:1 (1.5\degree $\times$ 0.01\degree, or 80 pc $\times$ 0.5 pc 
at a kinematic distance of 3.1 kpc).  Maps of \hnc emission, a tracer of dense molecular gas, 
made with the Australia Telescope National Facility Mopra telescope, 
show an excellent morphological match to the mid-IR extinction.  Moreover,
because the molecular line emission from the entire nebula has the same radial 
velocity to within $\pm 3.4$ \kms, the nebula is a single, coherent cloud and 
not the chance alignment of multiple unrelated clouds along the line of sight.  

The Nessie Nebula contains a number of compact, dense molecular cores which
have a characteristic projected spacing of $\sim$ 4.5 pc along the filament.  The theory of 
gravitationally bound gaseous cylinders predicts the existence of such
cores, which, due to the ``sausage'' or ``varicose'' fluid instability, fragment
from the cylinder at a characteristic length scale.  If turbulent pressure
dominates over thermal pressure in Nessie, then the observed core
spacing matches theoretical predictions.  We speculate that the formation of
high-mass stars and massive star clusters arises 
from the fragmentation of filamentary 
IRDCs caused by the ``sausage'' fluid instability that leads 
to the formation of massive, dense molecular cores.  The filamentary molecular 
gas clouds often 
found near high-mass star-forming regions (e.g., Orion, NGC 6334, etc.) may represent 
a later stage of 
IRDC evolution.

\end{abstract}


\section{Introduction}
\label{Introduction}

When infrared dark clouds (IRDCs) were originally identified 
as dark extinction features against the Galactic background at mid-IR wavelengths 
\citep{Perault1996,Egan1998,Carey1998,Carey2000,Hennebelle2001}, it was
not clear what role, if any, they might play in the process of star formation.
Their high column densities ($\sim 10^{23}$--$10^{25}$ cm$^{-2}$), 
together with their cold temperatures ($<25$ K) suggest that if stars 
do form within IRDCs, then IRDCs must comprise
an early phase in the process; otherwise the embedded stars would
have heated their surrounding gas and dust.  
Recently, compact molecular cores within
IRDCs have been found to contain embedded young stars or protostars (e.g.,
\citealt{Chambers2009}). A
few of these embedded young stellar objects will evolve into high-mass stars
(e.g., \citealt{Beuther2005,Rathborne2005,Rathborne2006,Rathborne2007,
Simon2006a,Pillai2006,Beuther2007}).  This star-formation activity, 
along with the similarity
in size and mass between IRDCs and cluster-forming molecular clumps, has led
to the suggestion that IRDCs are the birthplaces of all high-mass stars 
and clusters (e.g., \citealt{Rathborne2006,Wyrowski2008}).

One striking feature of IRDCs is their filamentary shape, evident
in the images from mid-IR Galactic surveys such as 
{\it MSX} \citep{Price2001}, GLIMPSE \citep{Benjamin2003}, 
and MIPSGAL \citep{Carey2005}.
Similar filamentary structures are also found in familiar examples of associated high-mass star-forming molecular 
clouds, such as Orion (e.g., \citealt{Plume2000,Tatematsu2008,Bally1987}).  
Because these warmer clouds bear a striking resemblance to IRDCs, 
the warm, star-forming filamentary clouds may represent a later evolutionary stage in the life of an IRDC. Indeed,
filamentary clouds may play an important role in high-mass star formation 
because they allow for enhanced accretion rates onto cores \citep{Banerjee2008,Myers2009}.

An extreme example of a filamentary IRDC is the ``Nessie" Nebula centered 
near $(l,b) = (338.4^\circ,
-0.4^\circ)$.  Nessie is identified as a dark mid-IR 
extinction feature in the GLIMPSE 
and MIPSGAL surveys (Figure \ref{figure1}).  If the mid-IR extinction represents a single object, then this IRDC has
an unusually large aspect ratio of over 150:1 ($1.5 ^\circ \times 0.01 ^\circ$).  
Nessie contains a number of extinction maxima which in other IRDCs are typically
associated with pre-stellar and protostellar cores with 
masses of order 100 $M_\odot$ (e.g., \citealt{Rathborne2006}).   

If the formation of IRDCs, perhaps due to the passage
of a spiral shock, results in filaments, then these long,
approximately cylindrical
structures may produce the necessary 
conditions for the formation
of dense molecular cores, and ultimately of clusters and high-mass stars.
Indeed, the theory of 
gravitationally bound gaseous cylinders predicts the formation of such
dense cores, which due to the ``sausage" or ``varicose" fluid instability, fragment
from the cylinder at roughly regular intervals 
(e.g., \citealt{Chandrasekhar1953,Nagasawa1987,Inutsuka1992,Nakamura1993,Tomisaka1995,
Tomisaka1996}).
This instability provides an attractive physical mechanism for 
high-mass star formation
and cluster formation in filamentary IRDCs.  

In this {\it Letter} we report on molecular line mapping of the Nessie Nebula.
We compare our results with the theory of self-gravitating fluid cylinders and
show that the observations broadly conform to the expectations of the 
``sausage'' instability.  We speculate that this mechanism may be important 
in the formation of high-mass stars and star clusters.

\section{Observations}
\label{Observations}

Millimeter molecular line observations of the Nessie Nebula were 
taken on 2008 August with the Australia Telescope National Facility 
(ATNF) 22 m Mopra Telescope.  The backend consisted of the wideband MOPS correlator
configured to observe simultaneously 16 separate passbands covering the 
frequency range 86 to 93 GHz.  Each of these passbands was 137.5 MHz wide, 
with 4096 spectral channels, corresponding to a velocity
resolution of 0.11 \kms.  We chose individual passbands to include molecular
lines of interest.  The brightest of the detected lines was \hnc (90.66 GHz). Although we detected a number of fainter
molecular lines, these yield similar results and will be presented in a subsequent paper.  

The nebula was mapped in the ``on-the-fly'' mode, in which data are collected while the telescope scans
in a raster pattern.  The raster rows were chosen to be perpendicular to the Galactic plane in order to avoid
striping artifacts parallel to the nebula's long axis.  The spacing between rows was 12'', or about 
1/3 of the Mopra beamwidth of
36'' FWHM \citep{Ladd2005}.  Flux calibration with a noise diode and 
an ambient temperature load was
performed every 20 minutes.

Data reduction and analysis utilized the ATNF livedata and gridzilla programs.
The individual spectra were co-added and gridded onto a uniform 15'' grid.  Linear baselines were removed from 
line-free channels.    All spectra are presented on the
antenna temperature ($T_A^*$) scale.  To convert to main-beam brightness 
temperatures, one should
divide the antenna temperatures by the main-beam efficiency of 0.49  \citep{Ladd2005}.  System temperatures for the observations 
were $\sim 180 - 300$ K,
which yielded a typical rms noise of $T_A^* = 0.14$ K in each spectral channel.

\section{Results}
\label{Results}

\begin{figure}
\includegraphics[trim=0cm 0cm 0cm 0cm,angle=-90,width=.965\textwidth,clip=true]{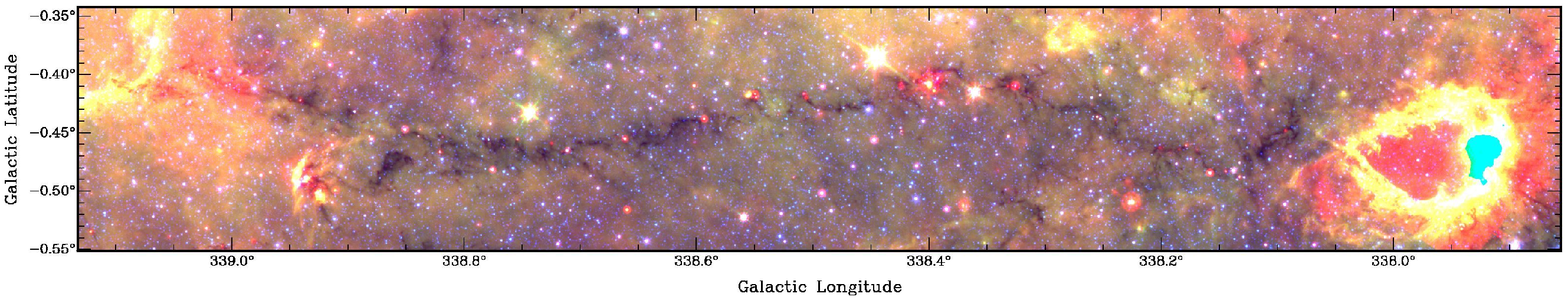}
\includegraphics[trim=.8cm 0cm .1cm 0cm,angle=-90,width=.965\textwidth,clip=true]{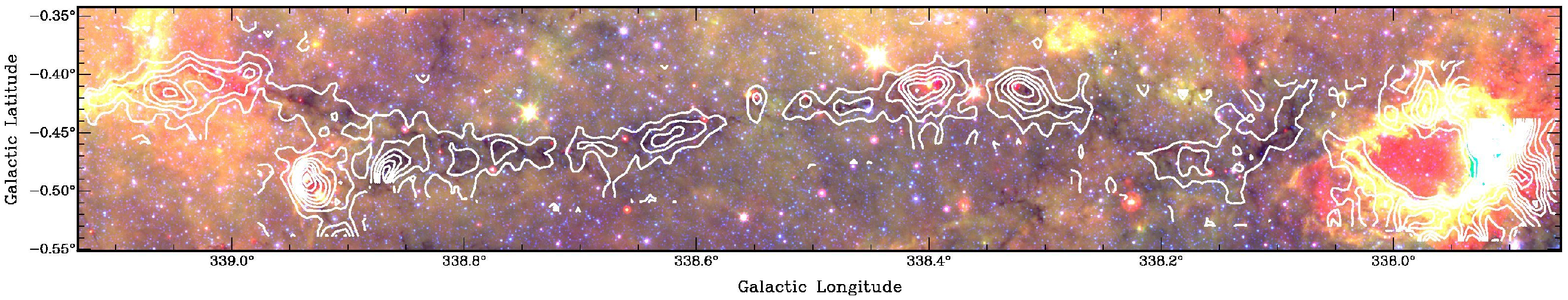}
\includegraphics[trim=1.4cm .25cm .4cm 0cm,angle=-90,width=1\textwidth,clip=true]{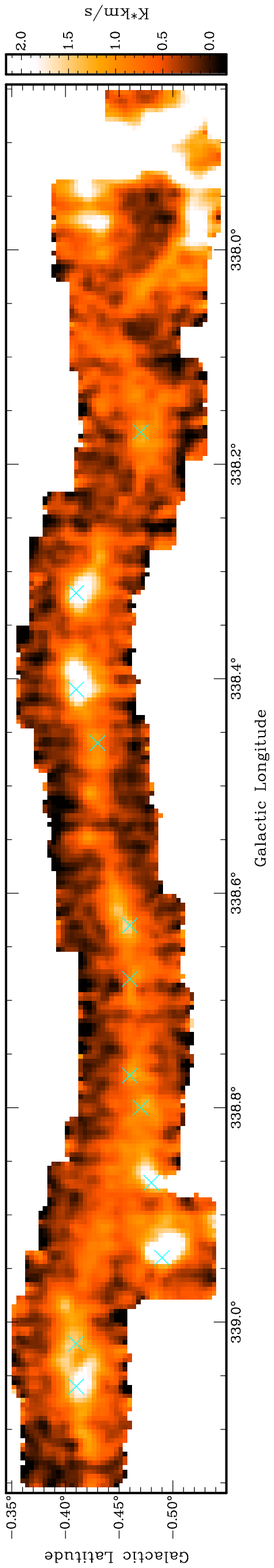}
\includegraphics[trim=1.6cm .25cm 0cm 0cm,angle=-90,width=1\textwidth,clip=true]{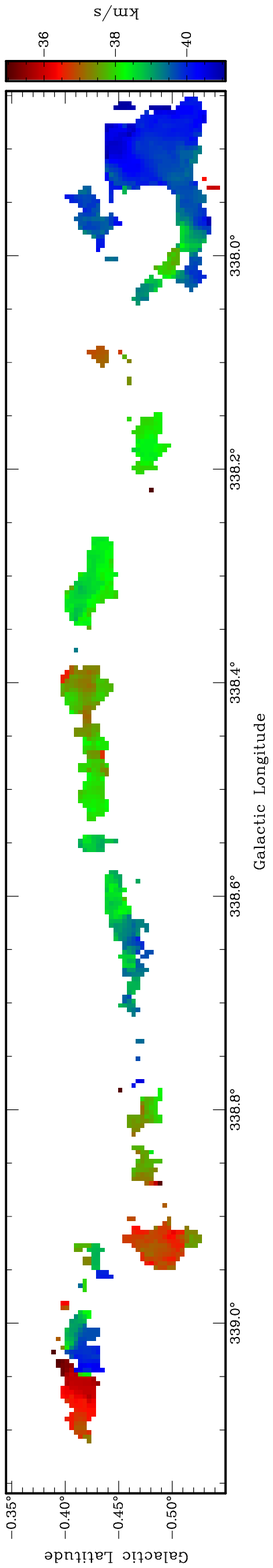}
\caption{Top panel: a false three-color image of the Nessie Nebula.  The 3.6 $\mu$m (blue) and 
8.0 $\mu$m (green) emission is from GLIMPSE, and the 24 $\mu$m (red) emission is from MIPSGAL.  
Second panel:  the false three-color mid-IR image of the Nessie Nebula from the top panel 
overlaid with integrated HNC (1-0) contours from the Mopra telescope.  
Note the excellent correspondence between the HNC emission and the 8 $\mu$m extinction.  
Third panel: an HNC (1-0) integrated intensity map from the Mopra telescope with core positions marked with cyan crosses. 
Bottom panel: a velocity-field (first moment) map from the HNC (1-0) map of the Nessie Nebula.  
All of the molecular emission is at the same velocity to within $\pm 3.4$ \kms, demonstrating that the filament is a single coherent object.}
\label{figure1}
\end{figure}

The Mopra integrated intensity \hnc maps  
closely correspond to the regions of mid-IR extinction (Figure \ref{figure1}).  The map consists of fainter, 
uniform emission associated with the filament, upon which compact, brighter regions we call ``cores'' are superposed.  
Gaussian fits to the \hnc lines demonstrate that every position within Nessie 
has essentially the same radial velocity, $-38$ $\pm 3.4$ \kms\ (Figure \ref{figure1}).  If the Clemens rotation curve is used, 
this velocity corresponds to a kinematic distance of 3.1 kpc \citep{Clemens1985}.

The molecular cores in Nessie are often associated star-formation activity.  
Specifically, many cores show the presence of excess 4.5 $\mu$m emission called ``green fuzzies,'' 
which indicates shocked gas (e.g., \citealt{Noriega-Crespo2004,Marston2004}), 
and bright 24 $\mu$m point sources, which indicate an embedded protostar (e.g., \citealt{Chambers2009}).  

The cores do not appear to be randomly spaced within the filament, but instead have a characteristic spacing.  The filament is thus reminiscent of a string with beads that are spaced sparsely but approximately uniformly along its length.  We used the clumpfind2d algorithm with a threshold of seven times the rms noise and 2.5$\sigma$ increments \citep{Williams1994} on the \hnc integrated intensity map 
to identify 12 molecular cores (see Figure \ref{figure1}).  The exact 
choice of clumpfind parameters will result in the identification of
slightly more or fewer cores, but the resulting core spacing is
quite insensitive to reasonable input parameters.  Since the cores are
relatively isolated and well separated, clumpfind works well and identifies the cores that one would tend to select by eye. 
The positions and selected properties of the cores are listed in Table \ref{coretable}.  Columns 1 and 2 give the Galactic coordinates of the position of peak integrated HNC intensity in each core, Column 3 the peak HNC integrated intensity, Columns 4 and 5 the angular size and physical extent, Column 6 the LTE mass (see Section \ref{Discussion}), and Column
7 whether or not the core contains a 24 $\mu$m point source.  The mean projected spacing between cores is roughly 4.5 pc.   

Although we prefer to use the molecular gas clumps
as our indicator of core spacing, since we thereby include pre-stellar
cores, another useful indicator of cluster-forming sites would be 24 $\mu$m point sources, which indicate embedded high-mass stars or protostars.  Indeed, about half of the molecular cores contain 24 $\mu$m
point sources (see Table \ref{coretable}).  For all of the unresolved 24 $\mu$m sources associated
with the nebula, we find typical spacing of 4.1 pc, a value within 10\% of the molecular core spacing.  Thus, both the spacing of gas clumps and
of 24 $\mu$m sources are in satisfactory agreement.

\section{Discussion}
\label{Discussion}

The existence of cores that are spaced at approximately a characteristic spacing within a highly filamentary cloud is 
in broad agreement with the predictions of fragmentation of a self-gravitating fluid cylinder 
due to the ``sausage" instability.  The theory's original presentation by
\citet{Chandrasekhar1953} treated an incompressible fluid.  Later
refinements included an isothermal, thermally supported cylinder with various 
magnetic field configurations \citep{Nagasawa1987,Inutsuka1992,Nakamura1993,Tomisaka1995,Tomisaka1996}.  

Although the fragmentation due to the ``sausage" instability of a self-gravitating 
fluid cylinder is similar to three-dimensional Jeans collapse, there are important differences.  
In spherical gravitational Jeans collapse, perturbations of all wavenumbers grow at the same rate, but
in cylindrical gravitational collapse perturbations with certain wavenumbers grow more quickly than others.  
Perturbations with the most unstable wavenumber will grow fastest,
and hence overdensities (cores) will tend to form at a length scale whose wavenumber 
corresponds to this fastest growing mode.  For a cylinder of infinite
radius, this length
scale is maintained even in the presence of a magnetic field parallel
to the filament's axis \citep{Nagasawa1987}.

The theory thus predicts (1) that multiple cores should form
within a filament, and (2) that the spacing between these cores should
be roughly periodic, with a
characteristic length scale equal to 
the wavelength of the fastest growing unstable mode 
of the fluid instability.  For an incompressible fluid,
this wavelength is $\lambda_{max} = 11 R$ where $R$ is the cylinder's radius
\citep{Chandrasekhar1953}.  In an infinite isothermal
gas cylinder, 
the relation is $\lambda_{max} = 22 H$, where $H$ is
the isothermal scale height given by $H = c_s(4 \pi G \rho_c)^{-1/2}$, where
$c_s$ is the sound speed, $G$ the gravitational constant, and
$\rho_c$ the gas mass density at the center of the filament ($R = 0$)
\citep{Nagasawa1987,Inutsuka1992}.  For isothermal cylinders
of finite radius $R$ surrounded by an external, uniform medium, the spacing
depends on the ratio between the cylinder radius and the isothermal 
scale height, $R/H$.  For $R \gg H$ the core
spacing approaches that for an infinite radius cylinder, $\lambda_{max} = 22 H$, 
but for $R \ll H$ the spacing reduces to that
of the incompressible case, $\lambda_{max} = 11 R$.  

We now compare the observations of Nessie to this theory under the assumption that
Nessie is well approximated by an isothermal cylinder.  If Nessie lies in the limit 
$R \ll H$, the theory reduces to that of an incompressible fluid,
and predicts cores spaced at a characteristic length scale $\sim 11R$.  
One can estimate the radius $R$ of Nessie from the radial extent of the sharp 
extinction edges evident in the GLIMPSE and MIPSGAL images.   This
suggests a radius of $R \sim 0.01 ^\circ$ or $\sim 0.5$ pc at the
kinematic distance of 3.1 kpc \citep{Clemens1985}.  Thus, 
one would expect cores to form
at a spacing of $\sim 5.5$ pc, in satisfactory
agreement with the observations
if the filament axis is mostly perpendicular to the line of sight.

If Nessie is treated as an isothermal gas cylinder with a central
volume density $n = 10^4$ cm$^{-3}$ and $T = 10$ K, appropriate for
cold, dense, molecular gas traced by \hnc emission, then the isothermal
scale height is $H = 0.05$ pc.  Thus, Nessie would be in the regime
$R \gg H$, and the theoretical spacing between
cores should be $\lambda_{max} = 22 H  \sim 1$ pc.  
This is smaller than the observed spacing
by about a factor of 5.  This discrepancy may arise from the fact
that turbulent pressure dominates over thermal pressure.

Theoretical studies of self-gravitating cylinders usually assume
that thermal pressure is the dominant gas pressure. 
However, because the observed linewidths in 
most molecular clouds typically
exceed the thermal sound speeds by large factors, turbulent pressure
usually dominates over thermal pressure.  If one replaces the
sound speed $c_s$ with the velocity dispersion $\sigma$ to account for
turbulent pressure \citep{Fiege2000}, the effective
scale height $H_{eff}$ is increased considerably over the thermal scale height $H$.  
Assuming a central volume density again of $n \sim 10^4$ cm$^{-3}$, and a FWHM linewidth of 2.5 \kms, 
typically observed toward Nessie, the effective
scale height is $H_{eff} \sim 0.17$ pc, which leads
to a spacing of $\lambda_{max} = 22 H_{eff} \sim 4$ pc between the cores. 
Again, the theoretical prediction is in satisfactory agreement with the 
observations, especially since the central density and hence the
scale height $H$ is uncertain.  Moreover, the theoretical assumption of a
uniform, isothermal cylinder (sometimes embedded in a uniform pressure
external medium) is obviously idealized, and the 
fragmentation length scale may well differ in more realistic treatments.

The theory also predicts that self-gravitating cylinders in equilibrium
have a
maximum, critical linear mass density (mass per unit length along
the cylinder's axis).  Above
the critical value, the cylinder would collapse radially into a line.
This critical
linear mass density
is given by $(M/l)_{max} = 2v^{2}/G$; where $v$ is the 
sound speed $c_s$ in the case of thermal 
support (Stodolkiewicz 1963, Ostriker 1964) and the turbulent 
velocity dispersion $\sigma$ in the case of turbulent support
\citep{Fiege2000}.
If turbulent pressure dominates over thermal pressure in Nessie, 
the critical linear mass density can be estimated as 
$(M/l)_{max} = 84 (\Delta V)^2$ $M_\odot$ $pc^{-1}$, 
where $\Delta V$ is in
units of \kms.  For typical line widths observed in 
Nessie ($\Delta
V \sim 2.5$ \kms), $(M/l)_{max} \sim 525$ M$_\odot$ pc$^{-1}$.  
[Note that magnetic fields
can alter the critical linear mass density, increasing
it for poloidal fields and decreasing it for toroidal fields, but probably
by small factors $\sim 1$ \citep{Fiege2000}.]

The mass per unit length in the Nessie Nebula can be estimated from
the \hnc fluxes, although this estimate is difficult due
to uncertainties in molecular abundances, depletion of molecules frozen onto
cold dust grains, and optical depth
effects.  Nevertheless, assuming optically thin \hnc emission,
LTE, a temperature of 10 K, and an HNC abundance 
of X$_{HNC/H_2} = 2 \times 10^{-9}$ \citep{Tennekes2006} appropriate
for cold clouds, we arrive at a typical linear mass density
in Nessie of 110 M$_\odot$ pc$^{-1}$.  Because this value is smaller than
the maximum predicted by theory, the estimated linear mass density 
is consistent with theoretical expectations.  
Unfortunately, the reported abundances
of HNC vary widely, with values for cold clouds ranging from 
$4.4 \times 10^{-12}$ to $2 \times 10^{-9}$ \citep{Wootten1978}.  These HNC abundance values give rise to 
a wide range in the linear mass density of up to $5 \times 10^4$ M$_\odot$ pc$^{-1}$.
Future observations of the dust continuum will better constrain the linear
mass density.  

It is interesting
to compare the value of $M/l$ obtained by theory and estimated in
Nessie with that measured for the molecular
filament associated with Orion, 
$M/l = 385$ M$_\odot$ pc$^{-1}$,
assuming M = $5 \times 10^3$ M$_\odot$ and $l = 13$ pc \citep{Bally1987}.
This value is within a factor of a few of that estimated
for Nessie, and also smaller than
the maximum value predicted by theory.

The maximum mass of the cores can also be estimated from theory.  The core 
masses should roughly be given by $M \sim \lambda_{max} (M/l)$.
Assuming the maximum, critical value for $M/l$ of $\sim 525$ M$_\odot$ pc$^{-1}$ 
for the case of 
turbulent pressure support, and a spacing of 4.5 pc, one expects the cores
to have maximum masses $M_{max} \sim 2400 M_\odot$.  If instead the average
linear mass density of 110 M$_\odot$ pc$^{-1}$ derived for Nessie is used,
the core masses should be no larger than about $500 M_\odot$. For Nessie, 
the estimated core masses based on the integrated \hnc fluxes over the 
solid angles range from $M \sim 30-390 M_\odot$ (Table \ref{coretable}). 

To summarize, if turbulent pressure, rather than thermal pressure, 
provides the support
for Nessie, the core spacing (4.5 pc) agrees with theoretical predictions.
Moreover, both the linear mass density and the core masses in Nessie
are also consistent
with the theory, since they are comparable to, but fall below, the maximum theoretical values.

We speculate that all
cluster formation arises from the fragmentation of filamentary 
IRDCs due to fluid instabilities that lead to massive, 
star-forming cores.  The periodic spacing of
such cores within Nessie strengthens this suggestion.  Moreover, filamentary molecular gas clouds are 
usually associated with high-mass star-forming regions, 
such as Orion, and these warmer clouds
bear a striking resemblance to IRDCs.  Such clouds may represent a later
evolutionary stage of a filamentary IRDC. 

\begin{figure}
\begin{center}
\includegraphics[angle=0,width=.8\textwidth,clip=true]{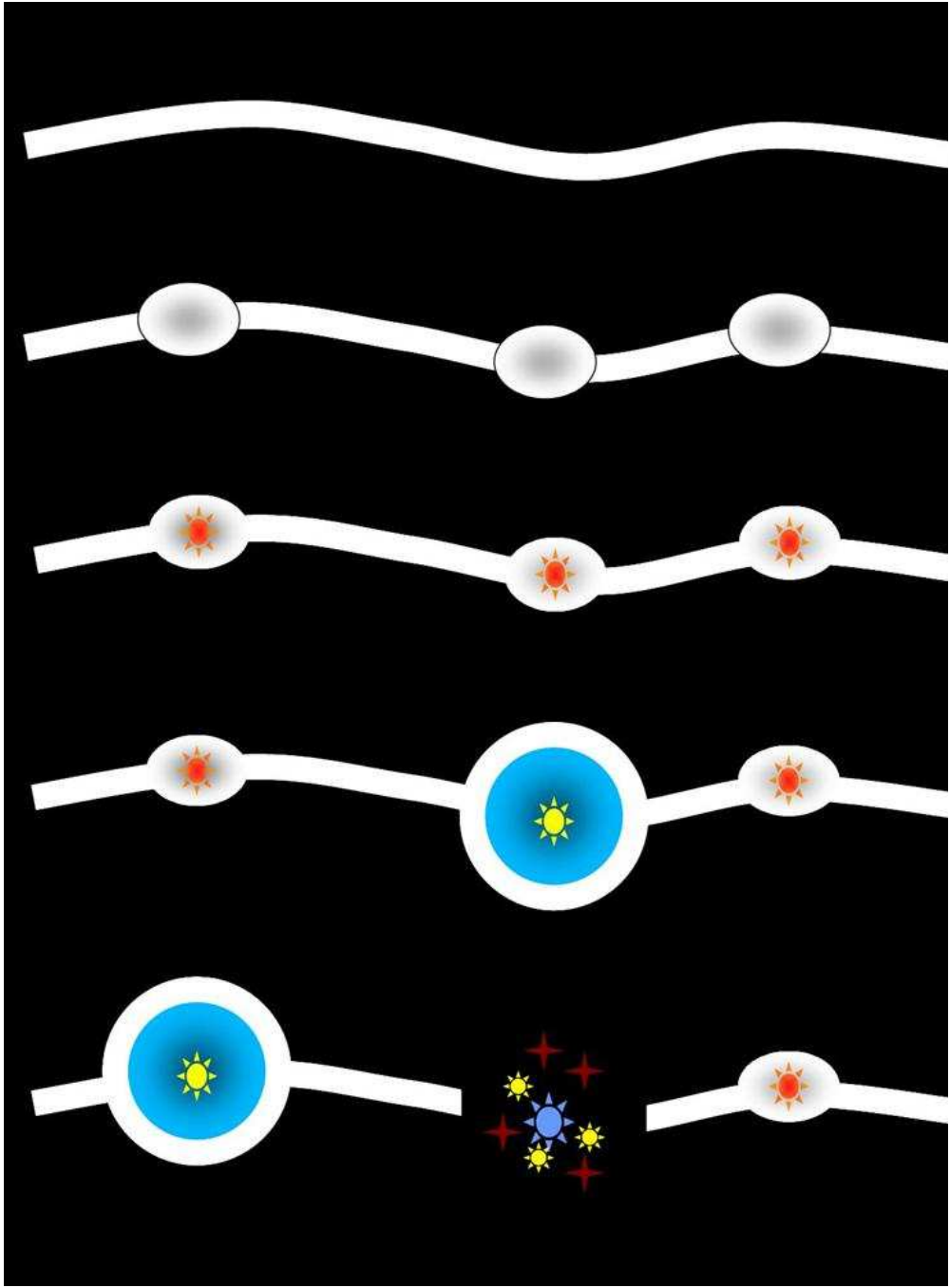}
\caption{Hypothetical sequence of IRDC evolution.  Time increases downward in the figure 
from top to bottom showing different evolutionary stages: a filamentary IRDC is formed, 
the filament fragments into dense cores, massive stars and \HII regions form within these cores, 
and finally star clusters emerge.}
\label{cartoon}
\end{center}
\end{figure}

A cartoon sketch of the proposed evolution of IRDCs is given in Figure \ref{cartoon}.  
Initially a filamentary IRDC is formed.  The filament reaches an equilibrium
configuration with a linear mass density close to the critical value.
The filament then fragments due to the
``sausage'' instability into roughly periodically spaced cores.  Such dense molecular
cores have sufficiently large masses to form clusters.  
The mass of the cores is set by a combination
of the turbulent velocity dispersion,
which determines the linear mass density, and the wavelength of the
fastest growing mode of the ``sausage'' instability. 
Thus, the cluster-forming
core mass, and ultimately, the mass of the resulting cluster,
is determined by the physics of the turbulent, self-gravitating cylindrical
fluid.  Within the more massive cores, high-mass protostars form.
When an embedded O or an early B protostar eventually reaches the main sequence, it
ionizes the surrounding gas and creates a bubble in the interstellar
medium \citep{Churchwell2006}.  Eventually the embedded cluster emerges.  

In the earliest phases, then, filamentary IRDCs with several cores
would be observed.  As the \HII regions form, bubbles would be associated
with the IRDCs, and eventually the IRDC would be segmented and disrupted
as the \HII regions grow and overlap.  Later, the filament would contain
one or more bright clusters, and appear similar to the filamentary molecular
clouds associated with high-mass star-forming regions such as 
Orion and NGC 6334 \citep{Kraemer1999}.  Examples of each of these proposed
stages can easily be found, from dark filamentary IRDCs like G$11.11-0.11$, sometimes
called the Snake \citep{Carey1998,Carey2000}, to IRDCs containing one
or more bubble-like \HII regions such as Nessie, to active high-mass 
star-forming regions like Orion and NGC 6334.

\section{Conclusions}
\label{Conclusions}

We have imaged a filamentary IRDC, the Nessie Nebula,
in \hnc molecular line emission with the ATNF Mopra Telescope.  Because
the \hnc radial velocities are uniform across the filament, Nessie is a coherent,
single cloud.  Both \hnc and mid-IR images reveal that Nessie
contains numerous dense cores, which appear to have a roughly periodic spacing
of 4.5 pc.  

The theory of self-gravitating, gaseous cylinders shows that 
they are unstable to a fluid
instability, called the ``sausage'' instability, which causes the filament to
fragment into cores with a
roughly constant spacing.  Thus, unlike the Jeans collapse of 
uniform media, the fragmentation of cylinders has a preferred length scale.  
Theory predicts that the fastest
growing wavelength scales as the radius $R$ or the isothermal scale height 
$H$ of the cylinder, depending on the ratio $R/H$.  In the conditions thought
to be appropriate to Nessie, the spacing between clumps is predicted to be
4 to 6 pc, in good agreement with the observed spacing of 4.5 pc.  

Because of the ubiquity of the association between filamentary molecular
clouds and high-mass
star forming regions, it is tempting to suggest that the physics of 
molecular filaments has a profound influence on cluster formation.  Indeed,
such filaments may be a necessary initial condition to form dense 
cluster-forming cores with the observed properties.  We speculate that the
``sausage'' instability may be the dominant physical mechanism 
to produce cores from filamentary IRDCs.  We suggest an evolutionary
sequence in which filamentary IRDCs fragment into cores, which in turn spawn star clusters and high-mass
stars.  Thus, familiar high-mass star-forming clouds such as Orion may have
begun their lives as filamentary IRDCs.

\acknowledgments
The authors gratefully acknowledge the funding support through NASA grant NAG5-10808 and 
NSF grants AST-0098562, AST-0507657, and AST-0808001.  The authors are grateful to Prof. Mark Johnson of Northwestern University for suggesting
the idea of varicose instabilities, to
Julia Duval-Roman for her assistance, and to an anonymous referee for important suggestions.

\clearpage

\clearpage

\begin{deluxetable}{cccccccc}
\tabletypesize{\scriptsize}
\tablecaption{Properties of Cores in the Nessie Nebula\label{coretable}}
\tablewidth{0pt}
\tablehead{
\multicolumn{2}{c}{Coordinates} &
\colhead{Peak Integrated HNC $T_A^*$} &
\colhead{Angular Size} &
\colhead{Physical Extent} &
\colhead{Mass} &
\colhead{24 $\mu$m Point Source?} \\
\colhead{\tiny{\it{l}\degr}} &
\colhead{\tiny{\it{b}\degr}} &
\colhead{(\it{K km s$^-1$})} &
\colhead{\tiny{(\it{arcmin$^2$})}} &
\colhead{\tiny{(\it{pc$^2$})}} &
\colhead{\tiny{(\msun)}} &
\colhead{\tiny{ }} }

\startdata

338.17	&	-0.47	&	1.15	&	3	&	2	&	80	& Y \\
338.32	&	-0.41	&	2.19	&	7	&	6	&	250	& Y \\
338.41	&	-0.41	&	2.26	&	9	&	7	&	320	& Y \\
338.46	&	-0.43	&	1.03	&	2	&	1	&	50	&   \\
338.63	&	-0.46	&	1.42	&	4	&	4	&	130	&   \\
338.68	&	-0.46	&	0.99	&	1	&	1	&	30	&   \\
338.77	&	-0.46	&	1.14	&	2	&	1	&	40	&   \\
338.8	&	-0.47	&	1.2	&	4	&	4	&	110	&   \\
338.87	&	-0.48	&	2.24	&	6	&	5	&	220	&   \\
338.94	&	-0.49	&	3.05	&	10	&	8	&	390	& Y \\
339.02	&	-0.41	&	1.47	&	7	&	6	&	220	&   \\
339.06	&	-0.41	&	2.02	&	9	&	7	&	310	& Y \\

\enddata

\end{deluxetable}

\end{document}